\documentclass[referee]{raa}            % referee version: for submission

\usepackage{graphicx,times}             %for PS/EPS graphics inclusion, new
\usepackage{natbib}
\usepackage{amssymb,amsmath}
\bibpunct{(}{)}{;}{a}{}{,}

\usepackage[a4paper=true,dvipdfm=true,pagebackref=true]{hyperref}
\hypersetup{colorlinks = true, linkcolor = green, anchorcolor = red, citecolor = blue, filecolor = red, pagecolor = red, urlcolor = red}

\begin{document}

   \title{The Jet of FSRQ PKS~1229$-$02 and its Misidentification as a $\gamma$-ray AGN}
   \volnopage{Vol.0 (20xx) No.0, 000--000}      %%preserved for Editor. DOn't remove!
   \setcounter{page}{1}          %%starting page, preserved for Editor. DOn't remove!
   \author{W.~Zhao
      \inst{1,2}
   \and X.-Y.~Hong
      \inst{1,2,3,4}
   \and T.~AN
      \inst{1,2}
  \and J.~YANG
      \inst{5}}

\institute{Shanghai Astronomical Observatory, 
   Chinese Academy of Sciences, 200030 Shanghai,
   P.R. China; {\it weizhao@shao.ac.cn}\\
        \and Key Laboratory of Radio Astronomy, Chinese Academy of Sciences, 210008 Nanjing, P.R. China\\
        \and University of Chinese Academy of Sciences, 19A Yuquanlu, Beijing 100049, People's Republic of China\\          
        \and Shanghai Tech University, 100 Haike Road, Pudong, Shanghai, 201210, People’s Republic of China\\
        \and Department of Earth and Space Sciences, Chalmers University of Technology, Onsala Space Observatory, SE-439 92 Onsala, Sweden\\
\vs\no
   {\small Received~~20xx month day; accepted~~20xx~~month day}}

\abstract{Flat-spectrum radio quasar PKS~1229$-$02 with a knotty and 
asymmetric radio morphology was identified as the optical and radio 
counterpart of a $\gamma$-ray source. In this paper, we study the 
properties, e.g. morphology, opacity, polarization and kinematics of 
the jet in PKS~1229$-$02 using radio interferometry. With our results,
we find that the knotty and asymmetric morphology of this source may 
probably shaped by the interaction between its anterograde jet and 
the nonuniform dense ambient medium. By reproducing a Spectral Energy 
Distribution of PKS~1229$-$02 with the obtained kinematic parameters, 
we find that the relativistic beaming effect in PKS~1229$-$02 is not
strong enough to produce the reported $\gamma$-ray emission, i.e. 
PKS~1229$-$02 may not be a $\gamma$-ray AGN. The misidentification
may probably due to the poor spatial resolution of the $\gamma$-ray 
detector of the previous generation.
\keywords{galaxies, individual, PKS~1229$-$02 --- 
galaxies, jets --- radio continuum, galaxies}
}

   \authorrunning{W.~Zhao et al. }            %author_head in even pages
   \titlerunning{The Jet of FSRQ PKS~1229$-$02}  % title_head in odd pages

   \maketitle

%
%________________________________________________ sections below
%
\section{Introduction}
\label{sect:Int}
Flat-spectrum radio quasar (FSRQ) PKS~1229$-$02 shows an asymmetric "core-jet-lobe" 
kpc-scale radio morphology. As shown in \cite{Kro1992}, radio emission extends 
for about 15~arcseconds towards northeast of the most luminous region 
(since this region has been identified as core by \citealt{Kro1992}, it will be referred 
as "core" hereafter), while only for about 5~arcseconds towards the southwest. 
The southwest jet often seems to be deflected as propagating outward: 
as shown by the 5 and 15~GHz VLA maps in \cite{Kro1992}, first it is deflected northward 
within 1 arcsecond from the core; further at about 2 arcseconds from the core, 
it is deflected southward and finally joints the southwest lobe.
The southwest kpc-scale jet is very knotty. It was also detected in the X-ray band \citep{Tav2007}, 
while some of its knots were also detected in the optical band \citep{Bru1997}. 
 
PKS~1229$-$02 was identified as the optical and radio counterpart of an EGRET-detected 
$\gamma$-ray source 3EG~J1230-0247 \citep{Tho1995, Har1999}, which is positionally 
associated with a Fermi-detected $\gamma$-ray source 3FGL~J1228.4-0317 \citep{AAA2009}. 
$\gamma$-ray sources are often identified as Active galactic nucleis (AGNs) with 
extremely relativistic jets aligned to our line of sight, e.g. FSRQ and BL lacs.  
Relativistic time dilation, Doppler boosting, together with projection effect, often bring 
about violent variations of luminosity, extremely high observed brightness temperature,  
and compact source structure in these $\gamma$-ray AGNs. In most theories propounded 
to explain the $\gamma$-ray emission from AGNs, relativistic beaming is always considered 
as a crucial role, no matter the emission particles are leptons \citep{BOT2000, MAR1985, MAR1992, 
BLO1996, DER1993, ARB2002, DON1995} or hadrons \citep{RAC2000, PRO1997, MUC2001} in these theories. 
But PKS~1229$-$02 only shows a very low optical variability \citep{Wil1992, Rai1998, Rom2002}
which is quite unusual for a $\gamma$-ray AGN, and whether there is extremely high observed 
brightness temperature or not, or whether its structure is compact or not, are still unknown. 

In this paper, we study the properties of the jet in PKS~1229$-$02, 
including the morphology, opacity, polarization, and kinematics. 
With the obtained results, we investigate the interaction 
between the jet and the surrounding medium, then we reproduce a 
Spectral Energy Distribution (SED) for PKS~1229$-$02 with the 
obtained kinematic parameters, and discuss the possibility for 
PKS~1229-02 as a $\gamma$-ray AGN  (or not). We use these 
cosmological parameters throughout this paper: 
$H_0=71$~km s$^{-1}$~Mpc$^{-1}$, $\Omega_{M}=0.27$, 
$\Omega_{\Lambda}=0.73$. At the distance of PKS~1229$-$02(z=1.045), 
1~mas corresponds to 8.126~pc, and 1~mas/yr corresponds to 28.5~$c$ \citep{Wri2006}.

\section{Observations and Data Reduction}
\label{SOD}

\begin{table}
\small \centering
\caption[]{Observations} \label{t1}
\tabcolsep 3mm
 \begin{tabular}{ccccccc}
  \hline\hline\noalign{\smallskip}
Obs.Code    &  Array      & Telescopes$~^{a}$ & Epoch   & Frequency         & Bandwidth  & Correlator \\
            &             &                 &         & (GHz)              & (MHz)     &            \\
   (1)      &  (2)        & (3)             & (4)     &  (5)          &  (6)      &  (7)        \\\hline\noalign{\smallskip}
\emph{VLA}  &             &                 &         &               &           &     \\
AH635       & VLA-C        & full array      & 1999.06  & 8.5, 22.5           & 50        & VLA \\
AH721       & VLA-A        & full array      & 2000.92  & 8.5, 22.5          & 50        & VLA \\\hline
\emph{VLBI} &             &                 &         &               &           &     \\
EH003       & EVN          & Ef Sh Jb Mc Nt Ht On Wb Ur Tr          & 1997.85  &C        & 28  &  MK III$~^{b}$       \\
BH065       & VLBA         & full array          & 2000.15  & 1.6        & 64              & VLBA$~^{c}$ \\
BH096       & VLBA         & full array          & 2002.55  & 4.9, 8.4, 15.0 & 64              & VLBA$~^{c}$   \\\hline
\emph{archived VLBI data} &             &                 &         &               &           &     \\
BB023       & VLBA         & full array          & 1997.35  &8.4        & 32  & VLBA$^{c}$\\
BL170       & VLBA         & BR FD HN KP LA MK NL OV SC & 2010.51 & 8.4 & 64 & VLBA$^{c}$ \\ 
\noalign{\smallskip}\hline
\end{tabular}
\parbox{150mm}
{Note. ---- $^{a}$ The telescopes with strong fringes: Ef:
Effelsberg, Sh: Shanghai, Jb: Jodrell Bank, Mc: Medicina, Nt:
Noto, On:Onsala, Hh: Hartebeesthoek, Sm: Simiez, Ur: Urumqi, Wb:
WSRT, Tr: Torun, BR: Brewster, FD: Fort Davis, HN: Hancock, KP: Kitt Peak, LA: Los Alamos, MK: Mauna Kea, NL: North Liberty, OV: Owens Valley, SC: St. Croix;
$^{b}$ The MK III correlator at MPIfR (Bonn, Germany);
$^{c}$ The VLBA correlator at NRAO (Socorro, USA)} 
\end{table}

From 1997 to 2002, we conducted a series of observations with Very Large Array (VLA), 
European VLBI Network (EVN), and Very Long Baseline Array (VLBA) to make studies on 
a sample of $\gamma$-ray AGNs with radio counterparts, among which, PKS~1229$-$02 
was observed in five tracks, including one track using VLA C-array at 8.5 and 22.5~GHz, 
one track using VLA A-array at 8.5 and 22.5~GHz, one EVN track at 5.0~GHz, one VLBA track 
at 1.6~GHz, and another VLBA track at 4.9, 8.4 and 15.0~GHz. 
All the VLA and VLBA tracks were conducted in dual circular-polarization 
mode while the EVN track was in left circular-polarization mode. 
The VLA and VLBA data are recorded in 2-bit format with a total bandwidth 
of 50 and 64~MHz respectively and correlated in Socorro, New Mexico, USA, 
while the EVN data were recorded in 2-bit format with a bandwidth of 28~MHz 
and correlated in Bonn, Germany. In additional, we acquired two sets of 
open-accessed 8.4~GHz VLBA data to study the kinematics of 
pc-scale jet of PKS~1229$-$02. Details of the observations are listed in Table \ref{t1}.

\begin{table}
\caption[]{VLA Images} \label{t2}
\tabcolsep 1mm
\begin{tabular}{ccccccc}
     \hline\hline\noalign{\smallskip}
   Epochs & Array  &  Frequency & Beam Size$~^a$               & $I_\mathrm{peak}~^b$  & $1\sigma~^c$       &Contours$~^d$\\
  (yr)    &        &  (GHz)     & (arcsec$\times$arcsec, degree)  &(Jy~beam$^{-1}$)    &(mJy~beam$^{-1}$)&(mJy~beam$^{-1}$)\\
  (1)     & (2)    &  (3)       & (4)                            &(5)                 &(6)       &(7)\\\hline\noalign{\smallskip}
         1999.06  &VLA-C    &  8.5       & 3.22$\times$2.34, 13.9       & 0.75         & 0.06     &$0.25\times(-1,1,2,4,...,2048)$    \\
                  &         &  22.5      & 1.35$\times$0.96, 17.4       & 0.51         & 0.17     &$0.66\times(-1,1,2,4,...,512)$    \\
         2000.92  &VLA-A    &  8.5       & 0.39$\times$0.25, 37.8       & 0.47         & 0.11     &$0.45\times(-1,1,2,4,...,1024)$     \\
                  &         &  22.5      & 0.17$\times$0.12, 36.5       & 0.15         & 0.22     &$0.86\times(-1,1,2,4,...,128)$     \\\hline\noalign{\smallskip}
\end{tabular}\\
Notes:\\
$^a$ The restoring beam FWHM size of the CLEANed images.  \\
$^b$ The peak flux density on the CLEANed images.\\
$^c$ The root-mean-square noise of the CLEANed images.\\
$^d$ The levels of the contours of the CLEANed images.
\end{table}

\begin{table}
\caption[]{VLBI Images} \label{t3}
\tabcolsep 1mm
\begin{tabular}{ccccccc}
     \hline\hline\noalign{\smallskip}
   Epoch &Array    &  Frequency  & Beam Size$~^a$    & $I_\mathrm{peak}~^b$  & $1\sigma~^c$  &Contours$~^d$\\
  (yr)   &         &  (GHz)      &(mas$\times$mas, degree) &(Jy~beam$^{-1}$)      &(mJy~beam$^{-1}$) &(mJy~beam$^{-1}$)\\
   (1)   &  (2)    & (3)         &(4)                      &(5)                   &(6)                 \\\hline\noalign{\smallskip}
         1997.85  &EVN      &  5         & 1.88$\times$1.36, 37.7       & 0.30               & 0.21            &$0.63\times(-1,1,2,4,...,256)$       \\
         2000.15  &VLBA     &  1.6       & 11.10$\times$4.94, -1.6      & 0.34               & 0.22            &$0.66\times(-1,1,2,4,...,512)$         \\
         2002.55  &VLBA     &  4.9       & 3.71$\times$1.69, -1.1       & 0.34               & 0.12            &$0.36\times(-1,1,2,4,...,512)$         \\
                  &         &  8.4       & 2.24$\times$1.04, -0.0       & 0.32               & 0.14            &$0.42\times(-1,1,2,4,...,512)$         \\
                  &         &  15.0      & 1.33$\times$0.57, -5.0       & 0.29               & 0.29            &$0.87\times(-1,1,2,4,...,256)$          \\\hline
\end{tabular}\\
Notes:\\
$^a$ The restoring beam FWHM size of the CLEANed images.  \\
$^b$ The peak flux density on the CLEANed images.\\
$^c$ The root-mean-square noise of the CLEANed images.\\
$^d$ The levels of the contours of the CLEANed images.
\end{table}

We performed the initial calibration with the AIPS 
(Astronomical Image Processing System) software package \citep{Dia1995} 
following the standard procedures of VLA and VLBI data 
reduction. For VLBA tracks, we performed several certain 
steps to calibrate the polarization: we determined and 
removed the variations of the parallactic angles with a 
procedure VLBAPANG; we determined the R-L delay 
difference with another procedure VLBACPOL on a 
highly polarized source (e.g. DA193, 3C~279); we calibrated 
the instrumental polarization with AIPS task LPCAL by 
using scans on a radio source with a compact structure (e.g. DA193, OQ~208); 
the absolute polarization angle was corrected by comparing 
the apparent Electric Vector Position Angle (EVPA) of calibrators with the quasi-simultaneous 
measurements in the VLA/VLBA Polarization Calibration Page \footnote{http://www.vla.nrao.edu/astro/calib/polar}. 
After initial calibration, we split the data into 
single-source files and imported them into the Caltech
VLBI Program DIFMAP \citep{She1994} for self-calibration 
and imaging. We performed self-calibration/Imaging loops 
for multiple iterations to obtain images with high dynamical 
ranges. The parameters of the images, e.g. beam size, peak 
intensity, and root-mean-square noise ($\sigma$) are listed 
in Table \ref{t2} and \ref{t3}.

\section{Results}

\subsection{Structure of the jet}
\label{SSMJ}

\begin{figure}
\centering
\includegraphics[width=14.0cm, angle=0]{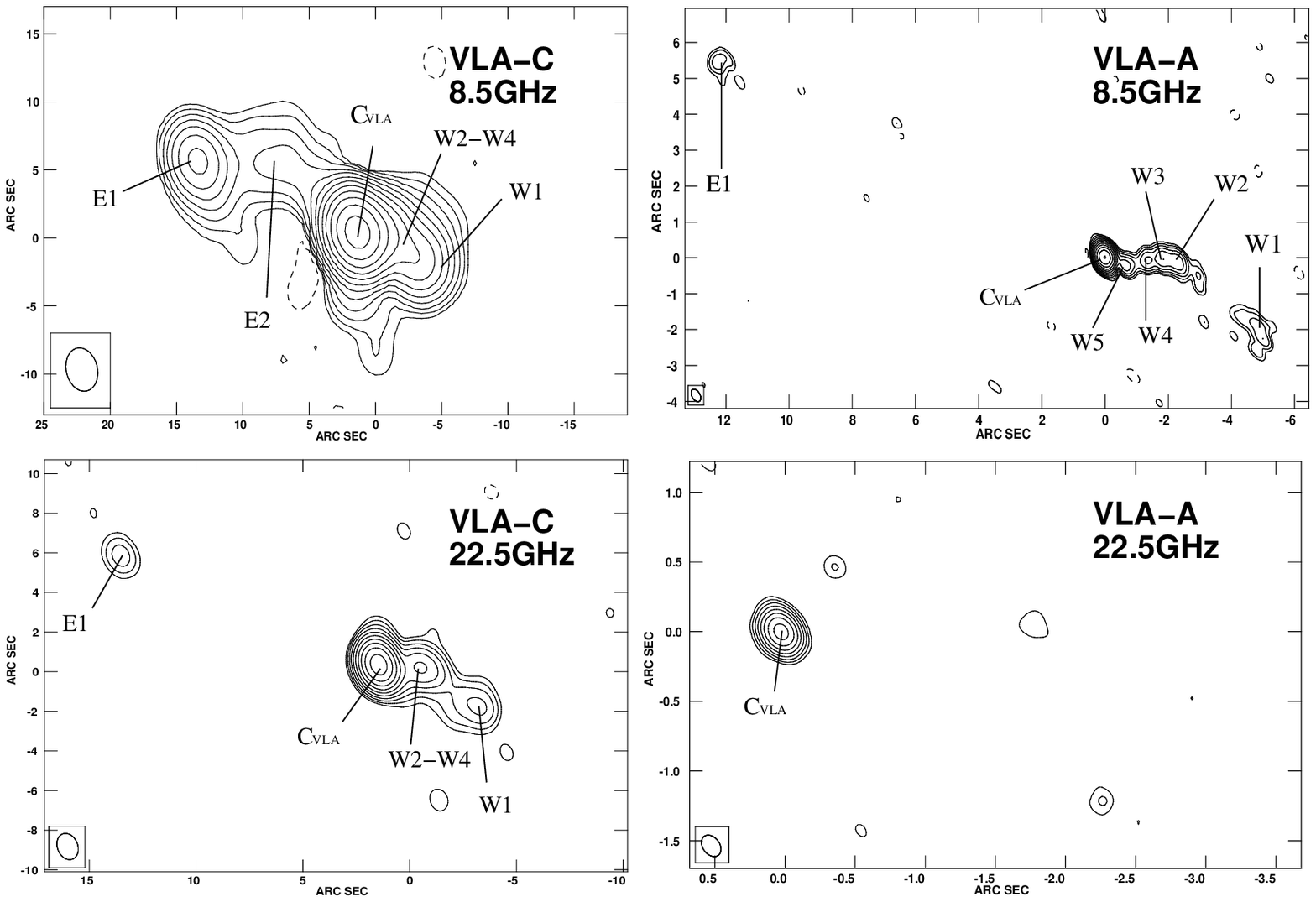}
\caption{VLA maps showing the kpc-scale morphology of PKS~1229$-$02.
The locations of the fitted Gaussians $\mathrm{C_{VLA}}$, 
W5, W4, W3, W2, W1, E2, and E1 are indicated on the maps.}\label{fVLA}
\end{figure}

\begin{figure}
\centering
\includegraphics[width=14.0cm, angle=0]{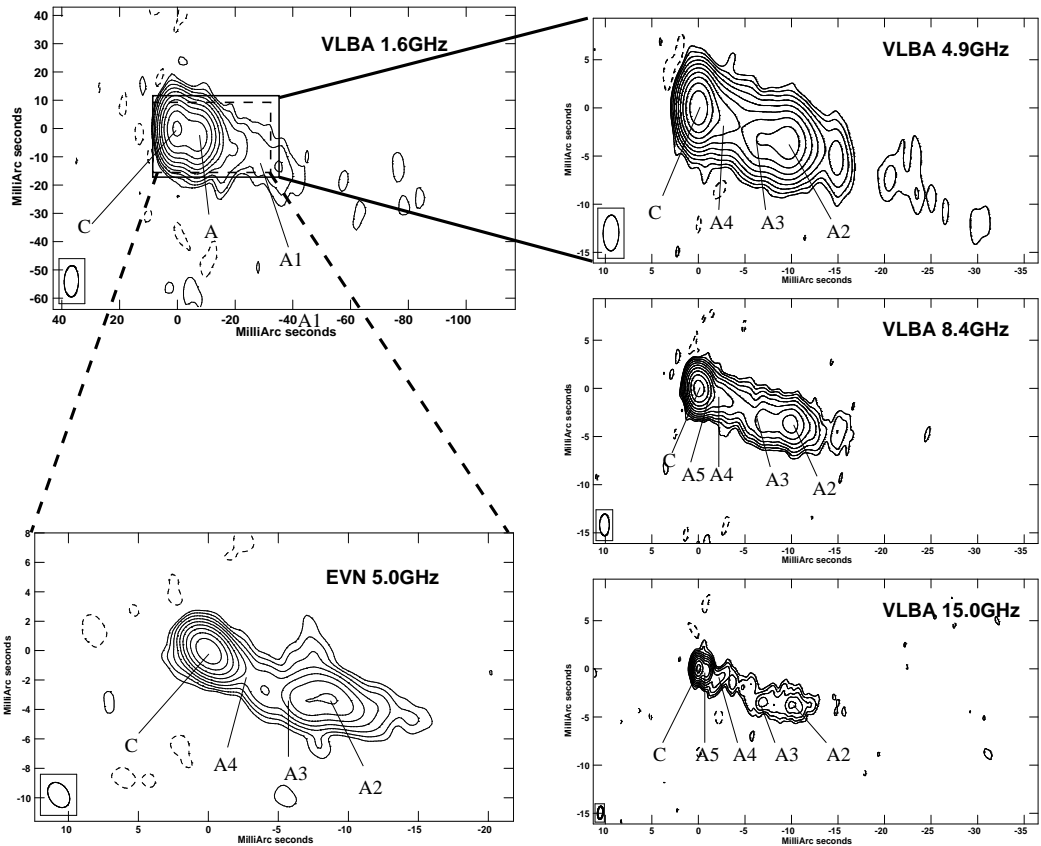}
\caption{VLBI maps showing the pc-scale morphology of PKS~1229$-$02.
The locations of the fitted Gaussians C, A5, A4, A3, 
A2 and A1 are indicated on the maps.}\label{fvlbi}
\end{figure}

We present the VLA and VLBI maps of PKS~1229$-$02 in Figure \ref{fVLA} 
and \ref{fvlbi} respectively (see details of the maps in Table \ref{t2} 
and \ref{t3}). As our VLA maps show, the kpc-scale morphology of 
PKS~1229$-$02 is quite consistent with that in \cite{Kro1992}. 
As our VLBI maps show, the pc-scale jet in PKS~1220$-$02 aligns well 
with the innermost part of the kpc-scale southwest jet in PKS~1220$-$02,
suggesting that the southwest jet goes toward our line of sight (referred as "the anterograde jet" 
hereafter), while its northeast counterpart goes against our line of sight (referred as 
"the retrograde jet" hereafter). The pc-scale jet is as knotty as its kpc-scale 
counterpart, but shows no significant curvature on the maps.

To further analyze the properties of the jet in PKS~1229$-$02, 
we fitted the $u-v$ data as Gaussian components with procedure 
MODELFIT in DIFMAP. The most luminous regions of the source on
both kpc- and pc-scales are often fitted with elliptical Gaussians, 
while the rest part of the source is often fitted with circular 
Gaussians. The uncertainties of the fitted parameters are 
estimated as in \cite{Foma2004} and \cite{Lee2008}. The fitted 
values are listed in Table \ref{t2} and \ref{t3}. The locations 
of the fitted Gaussians are indicated in Figure \ref{fVLA} and
\ref{fvlbi}. 

On kpc-scale, the core component which has already been identified 
by \cite{Kro1992} is renamed as $\mathrm{C_{VLA}}$, which is the 
only detected component on VLA A-array map at 22.5~GHz.
The retrograde lobe and jet are clearly detected on VLA C-array map at 8.5~GHz, 
and could be fitted with two Gaussian components named as E1 and E2 
respectively. But E2 is undetected on the maps of VLA C-array at 22.5~GHz 
and VLA A-array at 8.5~GHz. The knotty morphology of the anterograde 
jet is best revealed by VLA A-array map at 8.5~GHz, and it could be 
fitted with five Gaussian components named as W1 to W5 from the edge to 
the center of the source, in which W3, W4, and W5 may correspond to 
the radio counterparts of the optical knots detected by HST \citep{Bru1997}.
On the maps of VLA C-array, for the lower angular resolution, W5 
could not be resolved from the core. Neither W2, W3, and W4 could 
be resolved from each other, so we fit them as a Gaussian 
component named as W2-W4.

On kpc-scale, the asymmetric structure of PKS~1229$-$02 is clearly revealed by our results, 
as the integral flux density of the anterograde jet is a few times higher than that of the 
retrograde jet. The source is edge-brightened on kpc-scale, i.e. the jet components at 
the far-end of the source (e.g. W1, W2, E1) tend to have higher integral flux density than 
the jet components at the middle of the source (e.g. W3, W4, W5, E2).

On pc-scale, the most luminous component is fitted with a Gaussian 
named as C (referred as "VLBI core" hereafter), while the rest 
part of the source is fitted with five Gaussian components named as 
A1 to A5 from the edge to the center (A5 could not be resolved from the 
VLBI core by the observations under 8.4~GHz). These components align well 
with each other in a line with a position angle ($PA$) of $-113^{\circ}$ 
to the north, so we use the line with $PA\sim-113^{\circ}$ to the north 
as the approximate jet axis in the following analysis. 

To make clear if PKS 1229$-$02 is affected by major Doppler boosting or not,
we estimate the brightness temperature $T_\mathrm{b}$ for the 
pc-scale components in the parent-galaxy rest frame with the 
method in \cite{GUI1996}. 
\begin{equation}
T_\mathrm{b}= 1.77\times10^{12} (1+z) \frac{S_{\mathrm{ob}}}{\nu_{\mathrm{ob}}^{2}\theta_{d}^{2}}\mathrm{K}
\end{equation}
in which $S_{\rm ob}$ is the flux density in Jy, 
$\nu_{\rm ob}$ is the observing frequency in GHz, 
and $\theta_{d}$ is the angular size of a Gaussian component in mas. 
In this paper, $\theta_{d} =\sqrt{ab}$, in which, $a$ and $b$ are 
the length of the major and minor axis of a Gaussian component respectively. 
The estimated values of $T_\mathrm{b}$ are listed in the last column of Table \ref{t6}. 
We compare $T_\mathrm{b}$ of the VLBI core with $T_\mathrm{eq}$,
the brightness temperature of a radio source in the equipartition 
between the energy of the radiating particles and the magnetic field,
which is often used as an upper limit of the intrinsic brightness temperature.
As the Eq. 4a in \cite{rea1994},
\begin{equation}
T_\mathrm{eq}=1.6\times10^{12}h^{-2/17}F(\alpha)^{-2}\left\{\frac{1-(1+z)^{-1/2}}{1+z}\right\}^{2/17}\times(1+z)^{(2\alpha-13)/17}S_\mathrm{op}^{1/17}(\nu_\mathrm{op}\times10^3)^{(1-2\alpha)/17}\mathrm{K}
\end{equation}
in which $\alpha$ is the spectral index with a typical value of -0.75, 
and $F(\alpha)$ is a factor with a typical value of 3.4, then $T_\mathrm{eq}\sim 0.5\times10^{11}\mathrm{K}$.
As we see, $T_\mathrm{b}$ of the VLBI core is at the order of magnitude of 10$^{10}$K or 10$^{11}$K, almost 
the same as the $T_\mathrm{eq}$, suggesting that PKS 1229$-$02 is not affected by major Doppler boosting.

On both kpc and pc-scale, the structure of PKS~1229$-$02 
could not be described as "very compact". E.g. on kpc-scale, 
at 8.5~GHz, the integral flux density of $\mathrm{C_{VLA}}$ 
is 0.76 and 0.63~Jy on the maps of VLA C-array and A-array 
respectively, which means almost 20$\%$ of its flux density 
is resolved by the longer baselines of the VLA-A array.
Moreover, on pc-scale, the VLBI core takes only 52$\%$ to 
64.1$\%$ of the total integral flux density of the source.

\begin{table}
\caption[]{Kpc-scale Jet Components} \label{t5}
\begin{center}
\begin{tabular}{cccccccccc}
\hline\hline
   Epochs &Array    &  Frequency & Comp.   & $S_{c}~^a$    & r$~^b$                & PA$~^c$             & a$~^d$      &a/b$~^e$    & $\Phi~^f$    \\
   (yr)   &       &  (GHz)     &           & (mJy)      & (arcsec)            & (degree)         &(arcsec)   &       &   (degree)        \\
  (1)   &  (2)     & (3)     &  (4)       & (5)     & ( 6)       &  (7)             &(8)             &(9)     &(10)         \\\hline\noalign{\smallskip}
         1999.06  &VLA-C    &  8.5       & $\mathrm{C_{VLA}}$      & 764$\pm$27 & 0                & --             & 0.45$\pm$0.01     & 0.5   & 50.2      \\
                  &         &            & W2-W4   & 74$\pm$10  & 2.12$\pm$0.01  & -90.1$\pm$0.3  & 0.42$\pm$0.02   & 1.0   &           \\
                  &         &            & W1      & 40$\pm$6   & 4.87$\pm$0.09  & -112.0$\pm$1.0 & 1.48$\pm$0.17  & 1.0   &           \\
                  &         &            & E2      & 13$\pm$6   & 7.80$\pm$1.14& 46.8$\pm$8.4   & 5.04$\pm$2.28  & 1.0   &           \\
                  &         &            & E1      & 24$\pm$5   & 13.09$\pm$0.08   & 66.3$\pm$0.4   & 1.10$\pm$0.17  & 1.0   &           \\\noalign{\smallskip}
                  &         &  22.5      & $\mathrm{C_{VLA}}$      & 530$\pm$30 & 0                & --             & 196.0$\pm$7.2   & 0.58  & 27.4     \\
                  &         &            & W2-W4   & 42$\pm$8   & 1.94$\pm$0.08  & -94.5$\pm$2.4  & 0.95$\pm$0.17   & 1.0   &      \\
                 &         &            & W1      & 18$\pm$6   & 4.91$\pm$0.23   & -114.5$\pm$2.7 & 1.41$\pm$0.46  & 1.0   &      \\
                  &         &            & E1      &  5$\pm$3   & 13.34$\pm$0.12  & 65.2$\pm$0.6   & 0.57$\pm$0.26   & 1.0   &      \\\hline \noalign{\smallskip}
         2000.92  & VLA-A   & 8.5        & $\mathrm{C_{VLA}}$      & 630$\pm$40 & 0                & --              & 214.5$\pm$9.8   & 0.4   & -41.3      \\
                  &         &            & W5      & 19$\pm$7   & 0.71$\pm$0.04   & -107.9$\pm$2.8  & 0.24$\pm$0.07   & 0.85  & -67.6       \\
                  &         &            & W4      & 16$\pm$6   & 1.36$\pm$0.06  & -93.6$\pm$2.4   & 0.33$\pm$0.12   & 0.29  & -56.8      \\
                  &         &            & W3      & 21$\pm$7   & 1.82$\pm$0.04  & -91.4$\pm$1.1   & 0.27$\pm$0.07   & 0.63  & -21.0      \\
                  &         &            & W2      & 33$\pm$10  & 2.27$\pm$0.06  & -93.2$\pm$1.6   & 0.45$\pm$0.12   & 0.62  & 66.9      \\
                  &         &            & W1      & 31$\pm$25  & 4.98$\pm$0.75 & -112.1$\pm$8.7  & 1.901$\pm$1.50  & 0.38  & 61.6      \\
                  &         &            & E1      & 12$\pm$ 8  & 13.32$\pm$0.17  & 65.9$\pm$0.7    & 0.56$\pm$0.35   & 0.58  & -68.6       \\\noalign{\smallskip}
                  &         &22.5        & $\mathrm{C_{VLA}}$      & 180$\pm$20 & 0                & --              & 0.07$\pm$0.01    & 0.76  & -27.3       \\\hline 
\end{tabular}\\
\end{center}
Notes:\\
$^a$ The integral flux density.\\
$^b$ The distance to the "core" component.\\
$^c$ The position angle to the "core" component.\\
$^d$ The length of the major axis.\\
$^e$ The ratio between the length of the major and minor axis.\\
$^f$ The position angle of the major axis.\\
\end{table}

\begin{table}
\caption[]{Pc-scale Jet Components} \label{t6}
\resizebox{150mm}{75mm}{
\begin{tabular}{ccccccccccc}
\hline\hline
  Epochs &Array    &  Frequency & Comp.   & $S_{c}~^a$ & r$~^b$      & PA$~^c$     & a$~^d$      &a/b$~^e$    & $\Phi~^f$  &$T_\mathrm{b}~^g$ \\
  (yr)   &         &  (GHz)     &         & (mJy)   & (mas)  & (degree) &(mas)   &      &  (degree)        & (K)  \\
  (1)   &  (2)     & (3)     &  (4)       & (5)     & ( 6)    &  (7)   &(8)     &(9)     &(10)   & (11) \\\hline
1997.35  & VLBA    & 8.4        & C       & 429$\pm$32 & 0      & --    &0.41$\pm$0.02    & 1.00 & --    & 1.28$\times$10$^{11}$  \\ 
                  &         &            & A5      & 112$\pm$18 & 0.76$\pm$0.02   & -109.1$\pm$1.5    &0.46$\pm$0.04    & 1.00 & --    & 2.72$\times$10$^{10}$ \\
                 &         &            & A4      &  39$\pm$14 & 2.96$\pm$0.36   & -110.8$\pm$7.2    &2.17$\pm$0.72    & 1.00 & --    & 4.25$\times$10$^{8}$ \\
                  &         &            & A3      &  72$\pm$18 & 7.06$\pm$0.24   & -115.4$\pm$2.0    &1.97$\pm$0.48    & 1.00 & --    & 9.55$\times$10$^{8}$\\
                  &         &            & A2      &  77$\pm$17 & 10.25$\pm$0.15  & -109.4$\pm$0.9    &1.55$\pm$0.31    & 1.00 &  --   & 1.65$\times$10$^{9}$\\\hline
         1997.85  & EVN     & 5.0        & C       & 366$\pm$30 & 0               & --             & 1.05$\pm$0.07   & 0.40  & 66.0    &1.20$\times$10$^{11}$       \\
                  &         &            & A4      &  45$\pm$17 & 3.09$\pm$0.48   &-118.0$\pm$8.9  & 2.69$\pm$0.96   & 1.00  & --      &9.09$\times$10$^{8}$      \\
                  &         &            & A3      &  73$\pm$15 & 7.48$\pm$0.13   &-116.2$\pm$1.0  & 1.34$\pm$0.26   & 1.00  & --      &5.86$\times$10$^{9}$        \\
                  &         &            & A2      &  88$\pm$19 & 10.17$\pm$0.24  &-109.4$\pm$1.4  & 2.27$\pm$0.48   & 1.00  & --      &2.46$\times$10$^{9}$\\\hline
         2000.15  & VLBA    & 1.6        & C       & 387$\pm$24 & 0               & --             & 3.11$\pm$0.14   & 0.49  & 76.8    &1.15$\times$10$^{11}$    \\
                  &         &            & A       & 216$\pm$19 &  7.35$\pm$0.24   & -110.9$\pm$1.9 & 6.82$\pm$0.48   & 0.10  & 74.3    &6.58$\times$10$^{10}$    \\
                  &         &            & A1      &   8$\pm$4  & 33.90$\pm$2.71  & -112.2$\pm$4.6 &12.32$\pm$5.43   & 0.44  & -3.1    &1.66$\times$10$^{ 8}$    \\\hline
         2002.55  & VLBA    &  4.9       & C       & 397$\pm$27  & 0               & --             & 1.01$\pm$0.05 & 1.00  & --  &  0.59$\times$10$^{11}$   \\
                  &         &            & A4      &  65$\pm$11   & 2.93$\pm$0.16   & -115.0$\pm$3.1 & 2.22$\pm$0.32   & 1.00  & --      &1.98 $\times$10$^{9}$     \\
                  &         &            & A3      &  83$\pm$12  & 7.29$\pm$0.14   & -115.1$\pm$1.1 & 2.08$\pm$0.28   & 1.00  & --      &2.89 $\times$10$^{9}$     \\
                 &         &            & A2      & 131$\pm$16  & 10.35$\pm$0.12  & -110.9$\pm$0.6 & 2.18$\pm$0.23   & 1.00  & --      &4.19 $\times$10$^{9}$     \\
                  &         &  8.4       & C       & 344$\pm$24  & 0               & --             & 0.66$\pm$0.03   & 0.30  & 77.0    &1.35$\times$10$^{11}$      \\
                  &         &            & A5      & 101$\pm$14  & 0.86$\pm$0.01   & -101.1$\pm$0.9 & 0.42$\pm$0.02   & 1.00  & --      &2.94$\times$10$^{10}$     \\
                  &         &            & A4      & 56$\pm$12   & 3.15$\pm$0.26   & -115.3$\pm$4.8 & 2.70$\pm$0.53   & 1.00  & --      &3.91$\times$10$^{8}$     \\
                  &         &            & A3      & 65$\pm$13   & 7.84$\pm$0.21   & -115.2$\pm$1.5 & 2.20$\pm$0.42   & 1.00  & --      &6.84$\times$10$^{8}$     \\
                  &         &            & A2      & 91$\pm$14   &10.69$\pm$0.13   & -110.4$\pm$0.7 & 1.83$\pm$0.27   & 1.00  & --      &1.40$\times$10$^{9}$     \\
                  &         & 15.0       & C       & 340$\pm$24  & 0               & --             & 0.33$\pm$0.02   & 0.64  & 80.4    &0.78$\times$10$^{11}$       \\
                  &         &            & A5      & 124$\pm$15  & 0.88$\pm$0.02   & -102.8$\pm$1.4 & 0.43$\pm$0.04   & 1.00  & --      &1.08 $\times$10$^{10}$    \\
.                 &         &            & A4      &  35$\pm$12  & 3.52$\pm$0.30   & -109.5$\pm$4.9 & 1.82$\pm$0.61   & 1.00   & --      &1.68 $\times$10$^{8}$    \\
                 &         &            & A3      &  44$\pm$15  & 7.89$\pm$0.31   & -115.3$\pm$2.3 & 1.88$\pm$0.62   & 1.00   & --      &2.00 $\times$10$^{8}$     \\
                  &         &            & A2      &  65$\pm$16  & 10.85$\pm$0.20  & -110.3$\pm$1.1 & 1.66$\pm$0.40   & 1.00   & --      &3.77 $\times$10$^{8}$     \\
         2010.51  & VLBA    & 8.4        & C       &410$\pm$32   & 0     &         & 0.35$\pm$0.02 & 0.83   & 1.68     & 2.03$\times$10$^{11}$ \\ 
                  &         &            & A5      & 48$\pm$11   & 1.05$\pm$0.05  & -98.7$\pm$2.6   & 0.63$\pm$0.10 & 1.00   & --         & 6.11$\times$10$^{9}$ \\ 
                  &         &            & A4      & 59$\pm$21   & 3.46$\pm$0.57  & -112.8$\pm$9.5  & 3.37$\pm$1.15 & 1.00   &--          & 2.68$\times$10$^{8}$ \\ 
                  &         &            & A3      & 58$\pm$20   & 8.85$\pm$0.42  & -114.5$\pm$2.7  & 2.54$\pm$0.85 & 1.00   & --         & 4.61$\times$10$^{8}$ \\ 
                  &         &            & A2      & 82$\pm$20   & 11.52$\pm$0.23 & -110.4$\pm$1.2  & 2.00$\pm$0.47 & 1.00   &--          & 1.05$\times$10$^{9}$ \\\hline  
\end{tabular} }\\
Note:\\
$^a$ The integral flux density.\\
$^b$ The distance to the core component.\\
$^c$ The position angle to the "core" component.\\
$^d$ The length of major axis.\\
$^e$ The ratio between the length of the major and minor axis.\\
$^f$ The position angle of the major axis.\\
$^g$ The brightness temperature.\\
\end{table}  
\clearpage

\subsection{Opacity of the jet}
\label{SSOJ}

\begin{figure}
\centering
\includegraphics[width=8.0cm, angle=0]{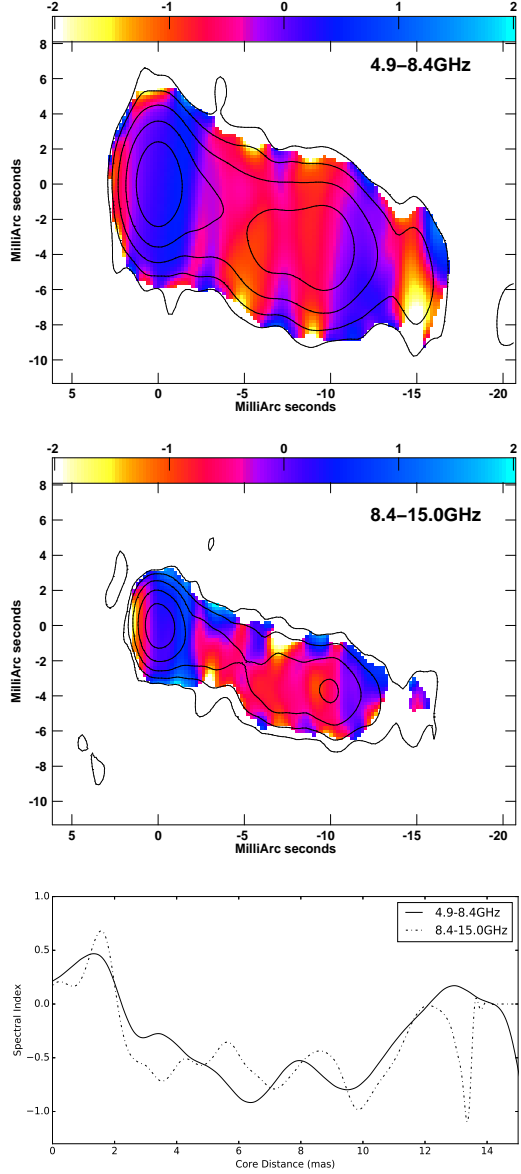}
\caption{The upper and middle panel shows the distribution 
of the spectral index calculated between 4.9 and 8.4~GHz 
and between 8.4 and 15.0~GHz respectively. The spectral 
index maps are produced with the data of the VLBA track 
conducted in the epoch of 2002.55. The pseudo-color are 
superimposed on the contour plots of total intensity of 
4.9 and 8.4~GHz respectively, in which the contours show 
$3\sigma\times$(1, 4, 16, 64, 256, 1024). The color is 
shown where the intensity is above $3\sigma$.
The lower panel shows the spectral index along the jet axis 
as a function of the core distance. }\label{fspec}
\end{figure}

To study the opacity of the kpc-scale jet in PKS~1229$-$02, 
we estimate the spectral index $\alpha$ of each jet components 
with the integral flux density fitted with VLA C-array data 
at 8.5 and 22.5~GHz, in the context of $S_{obs}\propto\nu_{obs}^{\alpha}$, 
in which $S_{obs}$ is the observed integral flux density, 
and $\nu_{obs}$ is the observing frequency. 
For the core component $\mathrm{C_{VLA}}$, $\alpha=-0.4$. 
For W2-W4 and W1, $\alpha=-0.6$ and -0.8 respectively. 
The flat spectrum of the anterograde jet suggests substantial 
low-frequency absorption, probably brought by the dense ambient 
medium. This is consistent with what was inferred 
by \cite{Kro1992}, that the anterograde jet interacts with the 
external environment as propagating outward. 
For component E2, $\alpha=-1.6$. Although E1 is not detected at 22.5~GHz, 
if we use $3\sigma$ of the 22.5~GHz map as the upper limit of its peak intensity, 
then the upper limit of $\alpha$ is -2.9. So the retrograde jet has an optical 
thin spectrum.

To study the opacity of the pc-scale jet in PKS~1229$-$02, we produce two 
spectral index maps with the data of the multi-frequency VLBA track conducted 
in the epoch of 2002.55. We first image the fully self-calibrated data of 
two adjacent frequencies (i.e. 4.9 and 8.4~GHz, 8.4 and 15.0~GHz) in the same $u-v$ range, 
then restore the resulting images with the same synthesized beam, 
and finally combine two images by aligning their map centers. 
The maps showing distribution of spectral index with pseudo-color 
are superimposed on the intensity contour plots of 4.9 and 8.4~GHz VLBA maps
respectively, and presented in the upper and middle panels of Figure \ref{fspec} respectively. 
The color is only shown where the intensity is above $3\sigma$.
We find that the inverted spectrum with $\alpha>0$ is detected 
in the vicinity of the VLBI core and the region between 10-14~mas 
from the map center. Between these two regions, steep spectrum 
with $\alpha\sim-1$ is detected. 

The lower panel of Figure \ref{fspec} shows $\alpha$ along the approximate 
jet axis evolving with the distance from the core (we use the position of the 8.4~GHz VLBI core, 
and core-shifts between frequencies are ignored). Within 2.0~mas from the core, inverted spectrum with $\alpha>0$ 
is detected, indicating that the emission from this region is very optical 
thick. Especially within 1.6~mas, spectral index increases as going outward, 
indicating that the opacity of the innermost jet is even higher than the VLBI core. 
Thus the free-free absorption (FFA) of the surrounding medium may attribute 
a lot to the high opacity as well as the synchrotron self-absorption (SSA) 
in the vicinity of the core. Beyond 2.0~mas, the spectrum goes steeper, 
i.e. the opacity decreases, as going outward. At 6-7~mas from the core, 
the spectral index reaches its minimum. Beyond the minimum, 
the spectral index increases as going outward, and the inverted spectrum 
appears again between 12 and 14~mas from the core.

\subsection{Kinematics of the pc-scale jet}
\label{SSKJ}
\begin{figure}
\centering
\includegraphics[width=10.0cm]{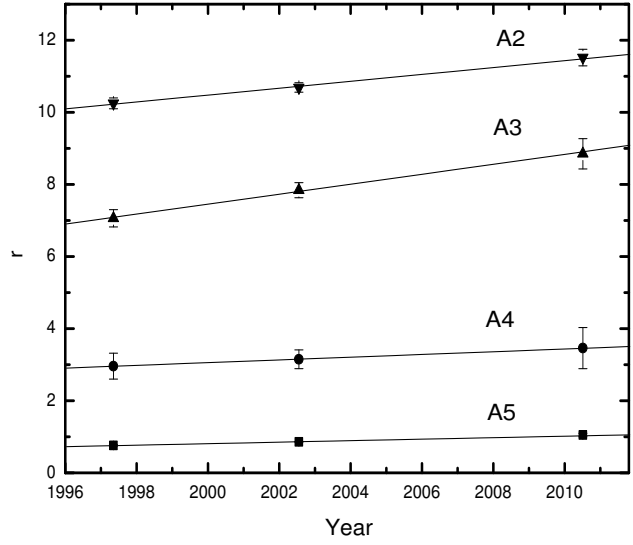}
\caption{Core distance of the pc-scale jet 
components A5, A4, A3, and A2 in 3 epochs 
(1997.35, 2002.55, and 2010.51) measured with 
8.4~GHz VLBA data, overlapping with the fitting 
results of their proper motion. }\label{fpro}
\end{figure}

With 3 epochs (1997.35, 2002.55, and 2010.51) of 8.4~GHz VLBA 
observations spanning 13 years, we obtain the proper motion 
of the pc-scale jet component A2, A3, A4 and A5. 
We fit their core distance increasing with time with a linear 
regression program, and use the inverse square of the uncertainty 
of the core distance as the weighting parameter in the fitting. 
Figures \ref{fpro} presents their core distance measured from 
3 epochs of 8.4~GHz VLBA data, overlapping with the fitting 
results of the proper motion.

The fitting proper motion for components A5, A4, A3 and A2 are 
0.021$\pm$0.002, 0.038$\pm$0.001, 0.138$\pm$0.009, and 0.096$\pm$0.007~mas~yr$^{-1}$ 
respectively, corresponding to the apparent transverse velocities ($\beta_{\rm app}$)
of $0.59\pm0.06$, $1.08\pm0.03$, $3.94\pm0.26$, and $2.72\pm0.20$ respectively in the unit of $c$. 
$\beta_{\rm app}$ increases from sub-luminal to superluminal between A5 and A3, 
and a mild deceleration follows in the region from A3 to A2.

As shown in \ref{SSMJ}, the average value of $T_\mathrm{b}$ of the VLBI core is about 
twice of the $T_\mathrm{eq}$, so it is proper to consider the Doppler factor $\delta\approx2$. 
Then the bulk Lorentz factor $\gamma$, the viewing angle $\theta$, and the intrinsic velocity $\beta$ in the unit 
of $c$ could be estimated as in \citet{Ghi1993}. 

\begin{equation}
\gamma= \frac{\beta_{\rm app}^{2}+\delta^{2}+1}{2\delta},
\end{equation}

\begin{equation}
\theta=\arctan~(\frac{2\beta_{\rm app}}{\beta_{\rm app}^{2}+\delta^{2}-1}),
\end{equation}

\begin{equation}
\beta=\frac{\beta_{\rm app}}{\sin\theta+\beta_{\rm app}\cos\theta},
\end{equation}

For A5, A4, A3 and A2, $\gamma$ is estimated to be 1.3, 1.5, 3.1, and 5.1 respectively, 
$\beta$ is 0.67, 0.76, 0.98 and 0.95 respectively, while $\theta$ 
is $19^\circ$, $27^\circ$, $23^\circ$, and $28^\circ$ respectively.

\subsection{Polarization of the pc-scale jet}
\label{SSPJ}
\begin{figure}
\centering
\includegraphics[width=16.0cm]{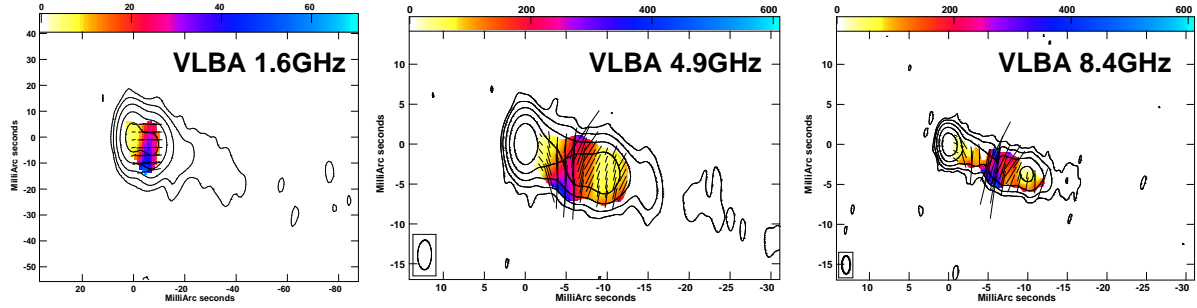}
\caption{Results of polarimetry of PKS~1229$-$02 at 1.6~GHz in 
the epoch of 2000.15, and at 4.9~GHz and 8.4~GHz in the epoch of
2002.55. The pseudo-color presents the fractional polarization $F_p$, 
and the unit used in the color-scale is Milli-percentage.
The vectors presents the EVPA. The pseudo-color and vectors are 
superimposed on the contour plots of total intensity, 
in which the contours show $3\sigma\times$(1, 4, 16, 64, 256).}\label{fpol}
\end{figure}

We successfully made polarimetry on PKS~1229$-$02 at 1.6~GHz in 
the epoch of 2000.15 and at 4.9 and 8.4~GHz in the epoch of 2002.55.
The results are shown in Figure \ref{fpol}, in which the fractional 
polarization ($F_p$) is presented in pseudo-color superimposed 
on the contour plots of total intensity, and the vectors represent 
the observed EVPA. The color is only shown where the polarized 
intensity is above $3\sigma$. 

As Figure \ref{fpol} shows, at 1.6~GHz, the core region is weakly 
polarized with $F_p<1\%$, while the jet within 10~mas from the 
core is polarized with $F_p<6\%$ and the EVPA is roughly 
parallel to the direction of the jet. 
At 4.9~GHz, the polarized emission is detected along the jet 
between 2 and 14~mas from the core; the highest value of $F_p$ 
($40\%$) appears at the edge of the jet, which is about 6~mas 
from the core. The EVPA is roughly parallel to the jet between 
2 and 3~mas from the core, while for the rest region, the EVPA 
is almost perpendicular to the jet. At 8.4~GHz, the configuration 
of the polarized emission is quite similar with that of 4.9~GHz. 
The polarized emission starts to be detected at only 1~mas from 
the core, and the peak value of the $F_p$ is even larger ($\sim60\%$). 

At 1.6~GHz, the core region is weakly polarized, while at 4.9 and 8.4~GHz 
it is not polarized at all. This is possibly due to the lower 
angular resolution at 1.6~GHz, thus the observed core is a mixture 
of the real core and the innermost jet.

\section{Discussion}

\subsection{Interaction between the jet and the ambient medium}

As we have mentioned in \ref{SSKJ}, between the jet component A5 and A3, 
the apparent transverse velocity of the pc-scale jet increases 
from sub-luminal to superluminal. Acceleration observed in pc-scale jets
in AGNs is usually interpreted as a consequence of the azimuthal 
magnetic field pressure gradient in the jet \citep{VK2004}. 
For PKS~1229$-$021, the acceleration seems very efficient, 
but magnetic field may not be the only factor affecting the 
bulk motion of its pc-scale jet. 

\begin{figure}
\centering
\includegraphics[width=12.0cm]{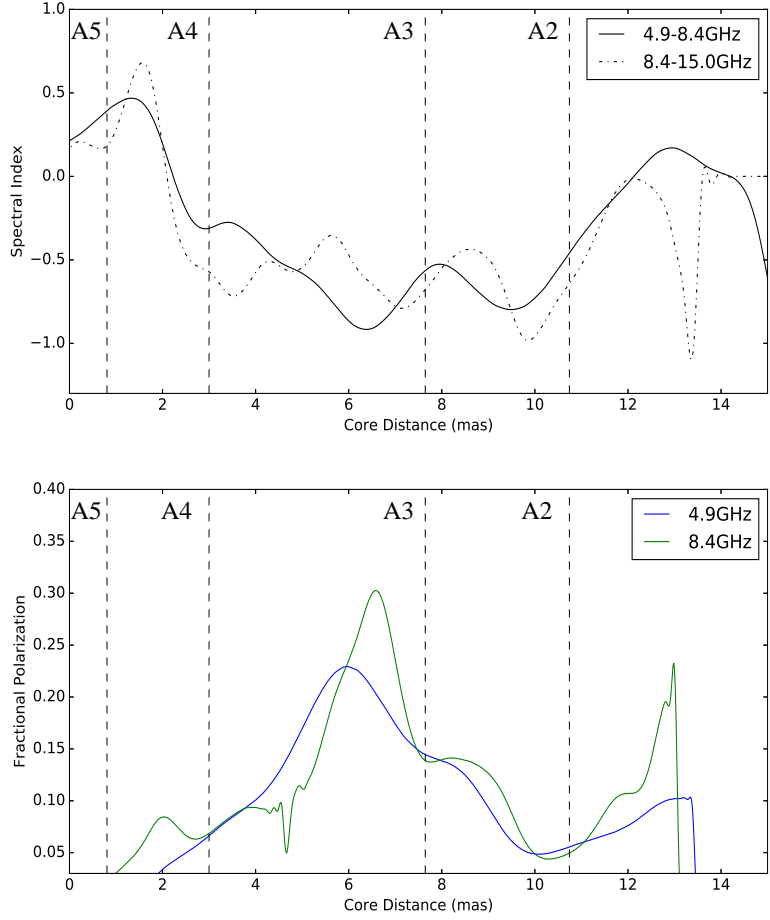}
\caption{The upper panel shows the spectral index  
evolving with the core distance as in the Figure\ref{fspec},
marked with fitted locations of the pc-scale jet 
components. The lower panel shows the fractional polarization 
$F_p$ evolving with the core distance, which is obtained
from the polarimetry at 4.9~GHz and 8.4~GHz in the epoch of 
2002.55 shown in Figure\ref{fpol}.}\label{fcom}
\end{figure}

We mark the locations of the pc-scale jet components on the figure 
showing the spectral index $\alpha$ evolving with the distance from 
the core (see the upper panel of Figure \ref{fcom}), and find there 
is a clear relation between the apparent transverse velocity and 
the opacity of the pc-scale jet in PKS~1229$-$021: the apparent 
transverse velocity increases as the opacity decreases. 
Moreover, the acceleration rate is not a constant, 
e.g. the acceleration rate is 0.21$c~\mathrm{mas^{-1}}$ from A5 to A4, 
much lower than that from A4 to A3, which is as high as 0.61$c~\mathrm{mas^{-1}}$.
As we have mentioned in \ref{SSOJ}, significant absorption features
are found between A5 and A4, indicating the surrounding medium might be 
very dense in this region, thus the interaction between the jet and 
medium may counteract a large part of the magnetic-driving acceleration.
From A4 to A3, the opacity decreases significantly, thus jet-medium interaction 
may get weaker in this region, so the magnetic-driving acceleration becomes 
much efficient and this is why acceleration rate is much higher in this 
region than that between A5 and A4.

A mild deceleration is found from A3 to A2, corresponding to 
the increasement of the opacity in the region, 
suggesting that the jet-medium interaction may have counteracted
the acceleration completely. Not only that, the jet is found 
brightened around A2, i.e. the integral flux density of the 
component A2 is higher than that of A3 and A4 in any epoch at 
any frequency. So, we infer that a shock is probably formed 
in this region, then the emitting particles are re-accelerated 
and the local magnetic field is enhanced. A part of the kinetic 
energy of the bulk motion is transfered to the radiant energy 
by the shock, thus the bulk motion slows down and the jet is 
brightened in this region. Another possible consequence of 
the enhancement of the local magnetic field is the increasing 
of the SSA opacity, so SSA may contribute a lot to the large
opacity beyond A2, as well as the FFA. 

The lower panel of Figure \ref{fcom} shows the factional polarization 
$F_p$ along the jet axis measured at 4.9 and 8.4~GHz (see \ref{SSPJ}) 
as a function of the core distance. Comparing with the upper panel, 
it is clearly seen that $F_p$ increases as the opacity decreases. 
The $F_p$ peaks between 6 and 7~mas from the core, just where the 
opacity approaches to its minimum. Beyond the peak, the $F_p$ decreases 
as the opacity increases until 10~mas from the core. It reminds us that, 
the surrounding medium might have significantly depolarized the emission.
Beyond 10~mas from the core, $F_p$ increases as the opacity increases 
as going outward, indicating that the shock formed around A2 might have 
enhanced the polarization and the SSA opacity at the same time. 
The surrounding medium may have also brought significant rotation of 
the EVPA, since for the region with high opacity, e.g. as Figure \ref{fpol}
shows, at 4.9 and 8.4~GHz, in the innermost part of the jet, the EVPA is parallel to jet, 
while for the outer region with lower opacity, the EVPA is perpendicular to the jet.

We conclude that, the anterograde jet of PKS~1229$-$02 is surrounded by 
nonuniform dense ambient medium, while the retrograde jet is not.
From the first few parsecs, to tens of kilo-parsecs away from its base, 
the anterograde jet always interacts intensively with the surrounding medium. 
The collisions between the jet and the medium have even evoked shocks,
building the knotty morphology of the anterograde jet from pc to kpc-scale.
The shocks have transfered a part of the kinetic energy of the jet to the 
radiant energy, and this is why the anterograde jet is brighter but more stubby than
its retrograde counterpart in kpc-scale.

\subsection{Misidentification as a $\gamma$-ray AGN}
 
As our observing results show, PKS~1229$-$02 is not compact on both kpc and pc-scales,
while the $T_\mathrm{b}$ of its VLBI core is as the same order of magnitude as the $T_\mathrm{eq}$. 
It shows a very low variability in the radio band as in the optical band, 
e.g. for all epochs of VLBI observations spanning 13 years in this paper, 
the total integral flux density of this source varies between a narrow 
range from 572 to 729~mJy. So we conclude that, in PKS~1229$-$02, the effects 
of relativistic time dilation, Doppler boosting, or projection are not prominent. 

Based on the results of our observations, we re-produce the SED of PKS~1229$-$02. 
It was once produced by \cite{Chi2003} and \cite{Tav2007} using  $\delta$ 
between 10 and 14, which is much larger than our estimated value ($\delta\approx2$). 
We use a model including synchrotron processes, self-Inverse Compton processes, 
outer-Inverse Compton processes, and thermal emission from the accretion disk. 
The details of the model are described in \cite{MAS2006, TRA2009, TRA2011}, 
the numerical code of the model is developed by Andrea Tramacere. 
For parameters like magnetic field, size of the jet, temperature of the accretion disk,
luminosity of the BLR, we use the same values as used in \cite{Tav2007}.
For the the Bulk Lorentz factor ($\gamma$) and the viewing angle ($\theta$) 
of the jet, based on the estimated value of pc-scale jet shown in \ref{SSKJ}, 
we adjust $\gamma$ between 1 and 6, while $\theta$ between 15$^{\circ}$ and 30$^{\circ}$ respectively
to fit our model to the historical data obtained from NASA/IPAC Extragalactic Database (NED) \footnote{http://ned.ipac.caltech.edu/}.

\begin{figure}
\centering
\includegraphics[width=15.0cm, angle=0]{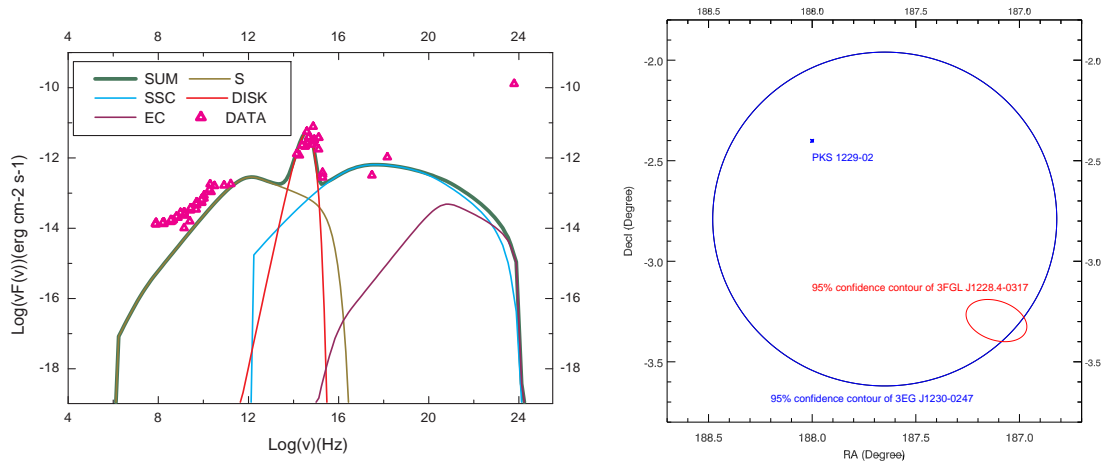}
\caption{The left panel shows the data of flux density of PKS~1229-02 obtained from NED (DATA),
overlapping with the best-fitting SED (SUM) using a model including 
synchrotron processes (S) , self-Inverse Compton processes (SSC), 
outer-Inverse Compton processes (EC), and thermal emission from 
the accretion disk (DISK). 
The right panel shows the positions of PKS~1229-02, 3EG J1230-0247, and 3FGL J1228.4-0317 on the sky plane.
The blue circle represents the 95\% confidence region of 3EG J1230-0247, 
the blue asterisk marks the optical position of PKS~1229$-$02, 
and the red elliptic presents the 95$\%$confidence 
region of 3FGL~J1228.4-0317.}\label{fsed}
\end{figure}

We find that if we exclude the $\gamma$-ray data, our model fits the rest 
data best when $\gamma=4$ and $\theta=18^{\circ}$ (i.e. $\delta=1.6$). 
The left panel of Figure \ref{fsed} presents the NED data overlapping 
with the best-fitting model. But, no matter how we adjust $\gamma$ 
and $\theta$, the prospected flux density in the $\gamma$-ray band 
is much lower than the NED value given by \cite{Tho1995, Har1999}. 
This result suggests that in PKS~1229$-$02 the relativistic beaming effect  
might not be strong enough to produce the reported $\gamma$-ray emission.
Although a small number of $\gamma$-ray sources are identified as AGNs without 
strong beaming effect, e.g. the radio galaxies M87, 
the compact steep-spectrum source 4C+39.23B,
or the narrow-line Seyfert 1 galaxy PKS 2004-447, 
PKS~1229$-$02 shows very little similarities with these sources.
Also we have compared the positions of PKS~1229$-$02, 3EG~J1230-0247 
and 3FGL~J1228.4-0317 on the sky plane. As the right panel of 
Figure \ref{fsed} shows, the optical position of PKS~1229$-$02 in Sloan Digital Sky Survey 
is right within the 95$\%$ confidence region of 3EG~J1230-024, but out 
of the 95\% confidence region of 3FGL~J1228.4-0317, and the distance 
from the quasar to the center of the confidence region is as far as $1.5^\circ$.
So PKS~1229$-$02 may probably could not produce any detectable $\gamma$-ray 
emission. It was misidentified as the radio and optical counterpart of a $\gamma$-ray source, 
due to the poor resolution of the detector of EGRET. 

\begin{acknowledgements}
The VLBA is an instrument of the National Radio Astronomy Observatory. 
The National Radio Astronomy Observatory 
is a facility of the National Science Foundation operated
under cooperative agreement by Associated Universities, Inc.
The European VLBI Network is a joint facility of European, 
Chinese, South African and other radio astronomy institutes funded by their national research councils.
\end{acknowledgements}

\end{document}